\documentclass[fleqn,usenatbib]{mnras}

\usepackage{graphicx}%
\usepackage{amsmath}%

\usepackage{newtxtext,newtxmath}
\usepackage[T1]{fontenc}

\DeclareRobustCommand{\VAN}[3]{#2}
\let\VANthebibliography\thebibliography
\def\thebibliography{\DeclareRobustCommand{\VAN}[3]{##3}\VANthebibliography}

\usepackage{textgreek}
\usepackage{xargs}
\usepackage{xspace}

\newcommand{\Halpha}{\text{H\textalpha}\xspace}
\newcommand{\Hbeta}{\text{H\textbeta}\xspace}
\newcommand{\Hgamma}{\text{H\textgamma}\xspace}
\newcommand{\Hdelta}{\text{H\textdelta}\xspace}

\newcommand{\Hepsilon}{\text{H\textepsilon}\xspace}

\newcommandx{\permittedEL}[6][1=O,2=III,3=,4=,5=,6=]{\text{{#1}\,{\sc {#2}}{#3}{#4}{#5}{#6}}\xspace}
\newcommandx{\semiforbiddenEL}[6][1=O,2=III,3=,4=,5=,6=]{\text{{#1}\,{\sc{#2}}]{#3}{#4}{#5}{#6}}\xspace}
\newcommandx{\forbiddenEL}[6][1=O,2=III,3=,4=,5=,6=]{\text{[{#1}\,{\sc{#2}}]{#3}{#4}{#5}{#6}}\xspace}

\newcommandx{\NVLr}[1][1=1243]{\permittedEL[N][v][\textlambda][#1]}
\newcommandx{\NVall}{\permittedEL[N][v][\textlambda][\textlambda][1239,][1243]}

\newcommandx{\CIIL}[1][1=1334]{\permittedEL[C][ii][\textlambda][#1]}

\newcommandx{\NIVL}[1][1=1486]{\semiforbiddenEL[N][iv][\textlambda][#1]}

\newcommandx{\CIVL}[1][1=1550]{\permittedEL[C][iv][\textlambda][#1]}

\newcommandx{\HeIIL}[1][1=1640]{\permittedEL[He][ii][\textlambda][#1]}

\newcommandx{\HeIIopL}[1][1=4686]{\permittedEL[He][ii][\textlambda][#1]}

\newcommandx{\OIIIL}[1][1=1666]{\semiforbiddenEL[O][iii][\textlambda][#1]}

\newcommandx{\NIIIL}[1][1=1750]{\semiforbiddenEL[N][iii][\textlambda][#1]}

\newcommandx{\NVL}[1][1=1239]{\semiforbiddenEL[N][v][\textlambda][#1]}

\newcommandx{\CIII}{\semiforbiddenEL[C][iii]}
\newcommandx{\CIIIL}[1][1=1909]{\semiforbiddenEL[C][iii][\textlambda][#1]}

\newcommandx{\NeIVL}[1][1=2424]{\forbiddenEL[Ne][iv][\textlambda][#1]}

\newcommandx{\MgIIL}[1][1=2803]{\permittedEL[Mg][ii][\textlambda][#1]}

\newcommandx{\NeVL}[1][1=3426]{\forbiddenEL[Ne][v][\textlambda][#1]}

\newcommandx{\OIIIaL}[1][1=4363]{\forbiddenEL[O][iii][\textlambda][#1]}

\newcommandx{\OIL}[1][1=6300]{\forbiddenEL[O][i][\textlambda][#1]}

\newcommandx{\OIIIopL}[1][1=5007]{\forbiddenEL[O][iii][\textlambda][#1]}
\newcommandx{\OIIIopLb}[1][1=4959]{\forbiddenEL[O][iii][\textlambda][#1]}
\newcommand{\OIIIopall}{\forbiddenEL[O][iii][\textlambda][\textlambda][4959,][5007]}

\newcommandx{\OIIL}[1][1=3727]{\forbiddenEL[O][ii][\textlambda][#1]}
\newcommandx{\OIILr}[1][1=3729]{\forbiddenEL[O][ii][\textlambda][#1]}

\newcommandx{\NeIIIL}[1][1=3869]{\forbiddenEL[Ne][iii][\textlambda][#1]}

\newcommandx{\CIIFIRL}{\forbiddenEL[C][ii][\textlambda][158\,\mum]}

\usepackage{booktabs}

\title[An offset AGN 740 million years after the Big Bang]{GA-NIFS: JWST discovers an offset AGN 740 million years after the Big Bang}

\author[H. {\"U}bler et al.]{\parbox{\textwidth}{
Hannah {\"U}bler,$^{1,2}$\thanks{hu215@cam.ac.uk}
Roberto Maiolino,$^{1,2}$
Pablo G.~P\'erez-Gonz\'alez,$^{3}$
Francesco D'Eugenio,$^{1,2}$
Michele Perna,$^{3}$
Mirko Curti,$^{4}$
Santiago Arribas,$^{3}$
Andrew Bunker,$^{5}$
Stefano Carniani,$^{6}$
St\'ephane Charlot,$^{7}$
Bruno Rodr\'iguez Del Pino,$^{3}$ 
William Baker,$^{1,2}$
Torsten B\"{o}ker,$^{8}$
Giovanni Cresci,$^{9}$
James Dunlop,$^{10}$
Norman~A.~Grogin,$^{11}$
Gareth C. Jones,$^{5}$
Nimisha Kumari,$^{12}$
Isabella Lamperti,$^{3}$
Nicolas Laporte,$^{1,2,13}$
Madeline A.~Marshall,$^{14,15}$
Giovanni Mazzolari,$^{16,17,1}$
Eleonora Parlanti,$^{7}$
Tim Rawle,$^{18}$
Jan Scholtz,$^{1,2}$
Giacomo Venturi,$^{7}$
Joris Witstok$^{1,2}$
}
\vspace{0.4cm}
\\
\parbox{\textwidth}{
$^{1}$Kavli Institute for Cosmology, University of Cambridge, Madingley Road, Cambridge CB3 0HA, UK\\
$^{2}$Cavendish Laboratory, University of Cambridge, 19 JJ Thomson Avenue, Cambridge, CB3 0HA, UK\\
$^{3}$Centro de Astrobiolog\'{\i}a (CAB), CSIC-INTA, Ctra. de Ajalvir km 4, Torrej\'on de Ardoz, E-28850, Madrid, Spain\\
$^{4}$European Southern Observatory, Karl-Schwarzschild-Stra\ss e 2, 85748, Garching, Germany\\
$^{5}$University of Oxford, Department of Physics, Denys Wilkinson Building, Keble Road, Oxford OX13RH, UK\\
$^{6}$Scuola Normale Superiore, Piazza dei Cavalieri 7, I-56126 Pisa, Italy\\
$^{7}$Sorbonne Universit\'e, CNRS, UMR 7095, Institut d' Astrophysique de Paris, 98 bis bd Arago, 75014 Paris, France\\
$^{8}$European Space Agency, c/o STScI, 3700 San Martin Drive, Baltimore, MD 21218, USA\\
$^{9}$INAF - Osservatorio Astrofisco di Arcetri, largo E. Fermi 5, 50127 Firenze, Italy\\
$^{10}$Institute for Astronomy, University of Edinburgh, Royal Observatory, Edinburgh EH9 3HJ, UK\\
$^{11}$Space Telescope Science Institute, 3700 San Martin Drive, Baltimore, MD 21218, USA\\
$^{12}$AURA for the European Space Agency, Space Telescope Science Institute, Baltimore, MD, USA\\
$^{13}$Aix-Marseille Universit\'e, CNRS, CNES, LAM (Laboratoire d'Astrophysique de Marseille), UMR 7326, 13388 Marseille, France\\
$^{14}$National Research Council of Canada, Herzberg Astronomy \& Astrophysics Research Centre, 5071 West Saanich Road, Victoria, BC V9E 2E7, Canada\\
$^{15}$ARC Centre of Excellence for All Sky Astrophysics in 3 Dimensions (ASTRO 3D), Australia\\
$^{16}$Dipartimento di Fisica e Astronomia, Universit`a di Bologna, Via Gobetti 93/2, I-40129 Bologna, Italy\\
$^{17}$INAF -- Osservatorio di Astrofisica e Scienza dello Spazio di Bologna, Via Gobetti 93/3, I-40129 Bologna, Italy\\
$^{18}$European Space Agency, ESAC, Villanueva de la Ca\~{n}ada, E-28692 Madrid, Spain\\
 }
}

\date{Accepted XXX. Received YYY; in original form ZZZ}

\pubyear{2023}

\begin{document}
\label{firstpage}
\pagerange{\pageref{firstpage}--\pageref{lastpage}}
\maketitle

\begin{abstract}
    A surprising finding of recent studies is the large number of Active Galactic Nuclei (AGN) associated with moderately massive black holes ($\rm \log(M_\bullet/M_\odot)\sim 6-8$), in the first billion years after the Big Bang ($z>5$).
    In this context, a relevant finding has been the  large fraction of candidate dual AGN, both at large separations (several kpc) and in close pairs (less than a kpc), likely in the process of merging.
    Frequent black hole merging may be a route for black hole growth in the early Universe; however, previous findings are still tentative and indirect.
    We present JWST/NIRSpec-IFU observations of a galaxy at $z=7.15$ in which we find evidence for a $\rm \log(M_\bullet/M_\odot)\sim7.7$ accreting black hole, as traced by a broad component of \Hbeta emission, associated with the Broad Line Region (BLR) around the black hole. This BLR is offset by 620~pc in projection from the centroid of strong rest-frame optical emission, with a velocity offset of $\sim$40~km/s. The latter region is also characterized by (narrow) nebular emission features typical of AGN, hence also likely hosting another accreting black hole, although obscured (type 2, narrow-line AGN). We exclude that the offset BLR is associated with Supernovae or massive stars, and we interpret these results as two black holes in the process of merging.
    This finding may be relevant for estimates of the rate and properties of gravitational wave signals from the early Universe that will be detected by future observatories like LISA.
\end{abstract}


\section{Introduction}

After one year of operations, one of the most surprising findings in the field of extragalactic astronomy produced by the James Webb Space Telescope \citep[JWST;][]{Gardner23, Rigby23} is the unexpectedly high abundance of active galactic nuclei (AGN) during the first billion years of cosmic history \citep[$z\gtrsim5.5$;][]{Harikane23, Maiolino23c, Greene23, Matthee23,Kocevski23,Yang23}. These are associated with accreting black holes and are identified through the properties of the spectral energy distribution \citep[SED; e.g.][]{Yang23}, through the presence of (narrow) nebular line emission typical of AGN \citep[e.g.][]{Uebler23,Maiolino23a,Scholtz23b}, and through the detection of a broad component of the Balmer lines, associated with the so-called Broad Line Region (BLR) of AGN \citep[e.g.][]{Kocevski23, Uebler23, Harikane23, Maiolino23c, Greene23, Matthee23}.
With the strongest lines of the hydrogen Balmer series, \Halpha and \Hbeta, being redshifted out of atmospheric windows for $z>4$ galaxies, many of these accreting black holes are now identified for the first time with JWST spectroscopy  \citep[primarily NIRSpec-MSA and NIRCam-slitless][]{Jakobsen22, Boeker23,Rieke2023_NC,Ferruit23} -- not least due to a remarkable lack of accompanying $X-$ray emission, previously taken as the main indicator to constrain the cosmic black hole accretion rate density (\citealp{Madau14}; but see also \citealp{Bogdan23, Goulding23}).

The JWST finding of the many AGN associated with moderately massive black holes has sparked the exploration of a large number of models, simulations and theories in order to explain their properties and abundances \citep[e.g.][]{Volonteri23,Trinca23,Schneider23,Pacucci23,Bennett23,Zhang23,Weller2023,DiMatteo23a,Mayer23}. These models involve various potential early black hole seeding scenarios, including: heavy seeds resulting from the direct collapse of primordial clouds; remnants of early massive stars; rapid merging of stars and (stellar mass) black holes in dense nuclear clusters. Episodes of super-Eddington accretion have also been suggested \citep{Schneider23,Trinca23,Bennett23}.

The role of merging between intermediate mass black holes has also been extensively investigated by models and simulations 
\citep{Barai2018,Barausse20,Volonteri20,Volonteri22,DiMatteo23a,DiMatteo23,valentini2021,Vito2022,Dimascia2021,Mannerkoski2022,Chen23}. This is not only a potential channel for the growth of black holes in the remote Universe, but also relevant for the detection of gravitational wave signals from early cosmic epochs with future gravitational waves experiments. 
Within this context, significant progress has been made observationally in recent years through the detection of dual quasars and AGN separated by a few/several kpc at high redshift (z$\sim$1--3) \citep{Mannucci22,Ciurlo2023,Perna23}, in a number significantly larger than predicted by cosmological simulations \citep[e.g.][]{RosasGuevara19, DeRosa19, Volonteri22, Chen23}. At even higher redshift (z$>$4), JWST multi-object spectroscopic data have suggested the presence of a significant population of dual AGN, on scales smaller than 1~kpc, based on the complex profile of their BLR, suggesting the presence of two separate BLRs associated with two accreting black holes \citep{Maiolino23c}. However, in the latter case no spatial information was available to confirm the presence of two AGN. Moreover, the complex profile of the broad lines may potentially also be associated with some peculiar properties of an individual BLR.

In this paper, we report the detection of a broad \Hbeta line in the $z=7.15$ galaxy system ZS7, which identifies it as a (type 1) AGN. ZS7 is also known by its CANDELS ID \citep{Grogin11, Koekemoer11, Nayyeri17} COSMOS13679 \citep{Pentericci16, Pentericci18}, COSY-0237620370 \citep{RobertsBorsani16, Laporte17, Witten23}, and COS-zs7-1 \citep{Stark17}. 
This galaxy was already identified to host an AGN, based on the detection of \NVL and \HeIIL in the rest-frame UV \citep{Laporte17}. However, the peculiarity of this system is that the BLR is offset from the primary (narrow) nebular line emitter. Additionally, also the latter emitter displays nebular line signatures typical of an AGN. We interpret our results in the context of supermassive black hole mergers, which will be detectable with future gravitational waves observatories like LISA.

Throughout this work, we use the AB magnitude system and assume a flat $\Lambda$CDM cosmology with $\Omega_m=0.315$ and $H_0=67.4$ km/s/Mpc \citep{Planck20}.  With this cosmology, $1''$ corresponds to a transverse distance of 5.27 proper kpc at $z=7.15$.

\section{NIRSpec-IFU Observations and Data Processing}\label{s:obs}

ZS7 was observed with the NIRSpec instrument \citep{Jakobsen22} in Integral Field Unit (IFU) mode \citep{Boeker22} on board JWST as part of the NIRSpec-IFU GTO programme `Galaxy Assembly with NIRSpec IFS' (GA-NIFS) under programme 1217 (PI: Nora L\"utzgendorf). The NIRSpec data were taken on May 9, 2023, with a medium cycling pattern of eight dithers and a total integration time of 4.1~h with the medium-resolution grating/filter pair G395M/F290LP, covering the wavelength range $2.87-5.27\mu$m (spectral resolution $R\sim700-1300$; \citealp{Jakobsen22}).
It was also imaged with the NIRCam instrument \citep{Rieke23} in eight filters though the PRIMER programme (PID: 1837, PI: James Dunlop).

\begin{figure*}
    \includegraphics[width=0.85\textwidth]{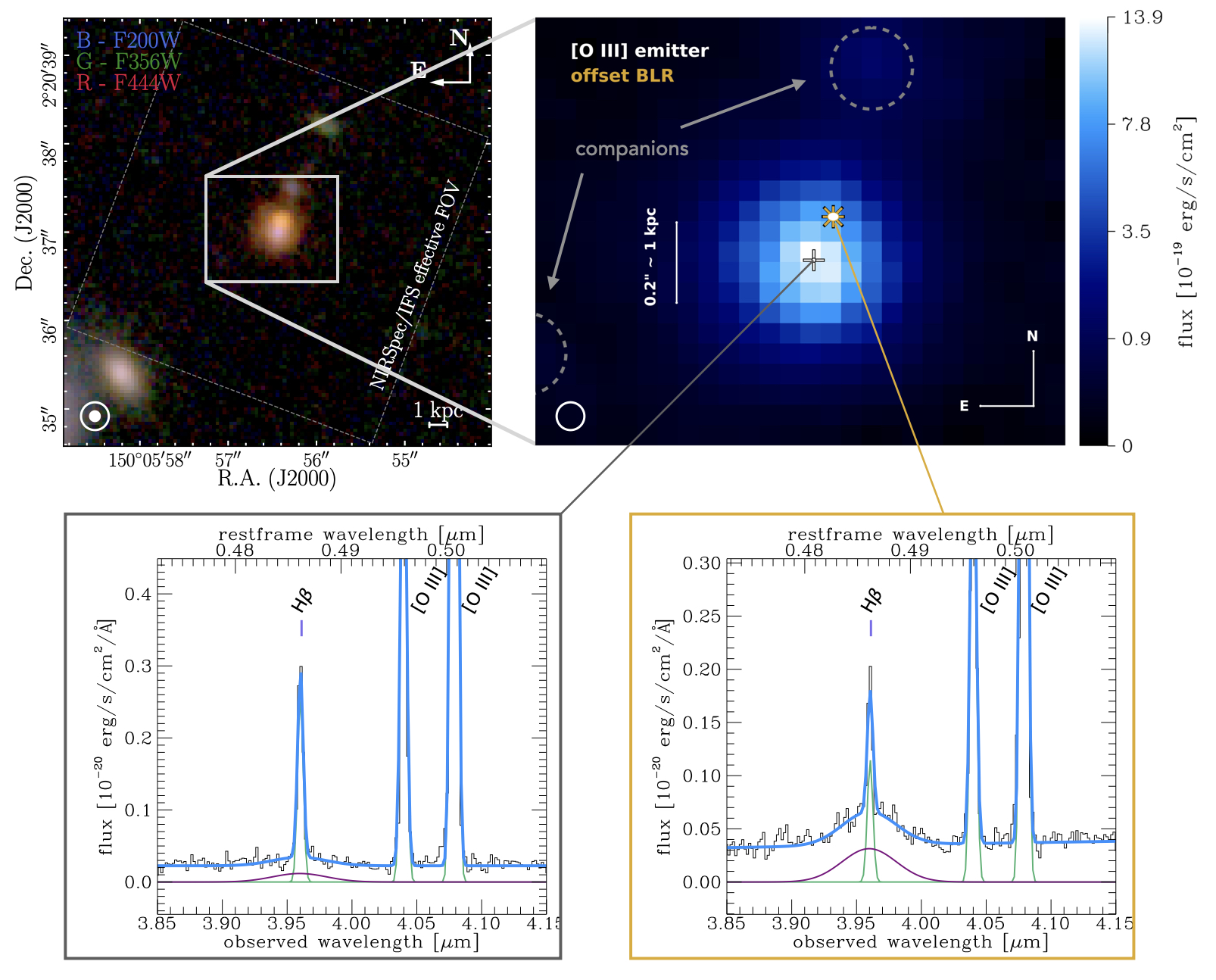}
    \caption{Top left: Color-composite image of ZS7 and its immediate environment created using public NIRCam data from the PRIMER program (PI: James Dunlop). The approximate NIRCam PSF is indicated by the white filled (F200W, $\sim0.06^{\prime\prime}$) and open (F444W, $\sim0.14^{\prime\prime}$) circles. The effective NIRSpec-IFU field of view (FOV) is indicated by the dashed white box.
    Top right: Zoom-in on the environment of ZS7, as indicated by the rectangular solid white box in the top left panel. The color scale shows a line map integrated over 90~\AA\, around the \OIIIopL position, where we indicate the positions of two faint companions detected spectrally at the same redshift with dashed grey circles. The centroids of the \OIIIopL emission and the BLR emission are indicated by a white plus and a golden star, respectively, where the small white circles on top of the symbols indicate five times the fitting errors on the centroid positions. The approximate NIRSpec PSF at $4\mu$m is indicated by the white circle.
    The centroid of the \OIIIopL emission coincides with the NIRCam F200W emission, while the BLR location coincides with the F444W emission, reflected in the color gradient in the NIRCam false-color image.
    Bottom: Zoom-in on the spectra extracted from a single spaxel at the \OIIIopL centroid position (left) and at the BLR location (right). We show our full fit in blue, with narrow emission lines in green and the BLR component in purple. 
    At the BLR location, the broad \Hbeta emission is evident, while the \OIIIopall emission is narrow at all locations. Some flux of the BLR point-source is spread through the point-spread function (PSF$_{\rm FWHM}\sim0.13-0.14''$ at $4\mu$m) and therefore is still visible at the \OIIIopL centroid position.}
    \label{f:target}
\end{figure*}

Raw data files were downloaded from the Barbara A.~Mikulski Archive for Space Telescopes (MAST) and subsequently processed with the {\it JWST} Science Calibration pipeline\footnote{\url{https://jwst-pipeline.readthedocs.io/en/stable/jwst/introduction.html}} version 1.8.2 under the Calibration Reference Data System (CRDS) context jwst\_1068.pmap. We made several modifications to the default reduction steps to increase data quality, which are described in detail by \cite{Perna23} and which we briefly summarise here. 
Count-rate frames were corrected for $1/f$ noise through a polynomial fit. 
During calibration in Stage 2, we removed regions affected by failed open MSA shutters. We also removed regions with strong cosmic ray residuals in several exposures.
Remaining outliers were flagged in individual exposures using an algorithm similar to {\sc lacosmic} \citep{vDokkum01}: we calculated the derivative of the count-rate maps along the dispersion direction, normalised it by the local flux (or by three times the rms noise, whichever was highest), and rejected the 96\textsuperscript{th} percentile of the resulting distribution \citep[see][for details]{DEugenio23}. 
The final cube was combined using the `drizzle' method, for which we used an official patch to correct a known bug.\footnote{\url{https://github.com/spacetelescope/jwst/pull/7306}} The main analysis in this paper is based on the combined cube with a pixel scale of $0.05''$.
We used spaxels away from the central source and free of emission features to perform a background subtraction.

\section{Data Analysis}\label{s:data_an}

\subsection{Spectral fitting and detection of broad \Hbeta}\label{s:fitting}

Fig.\ref{f:target} shows two spectra (zoomed around \Hbeta) extracted from the IFU cube at the location of the \OIIIopL centroid and at a location about 0.12$''$ to the NW, where a broad component of \Hbeta is clearly visible.
We fit the spectra with a combination of Gaussian functions for the line emission, and a power-law continuum. We include a narrow component for each of the following lines: \OIIL, \OIILr, \NeIIIL, He~I$\lambda3889$+H8, [Ne~III]$\lambda3967$, \Hepsilon, \Hdelta, \Hgamma, \OIIIaL, \Hbeta, \OIIIopLb, \OIIIopL. The intensity ratio of the [O~III] doublet is fixed to \OIIIopL/\OIIIopLb = 2.98, based on atomic physics \citep{Osterbrock06}. We note that the [O~II] doublet as well as [Ne~III]$\lambda3967$ and \Hepsilon are blended.
We show our full fits to spectra extracted from the BLR location and the centroid of the \OIIIopL emission in Appendix~\ref{a:fits}.

As mentioned, for \Hbeta, a broad component is observed in some spaxels. We see no evidence for broad components in other lines. 
As it will be discussed later on, we interpret this broad \Hbeta component as a BLR, offset relative to the \OIIIopL centroid.

Following \cite{Uebler23}, we re-scale the formal noise based on the `ERR' extension with a measurement of the standard deviation in regions free of line emission to take into account correlations due to the non-negligible size of the PSF relative to the spaxel size. This increases the noise by a factor of up to two for individual spaxels.

\begin{figure*}
    \centering
    \includegraphics[width=0.3\textwidth]{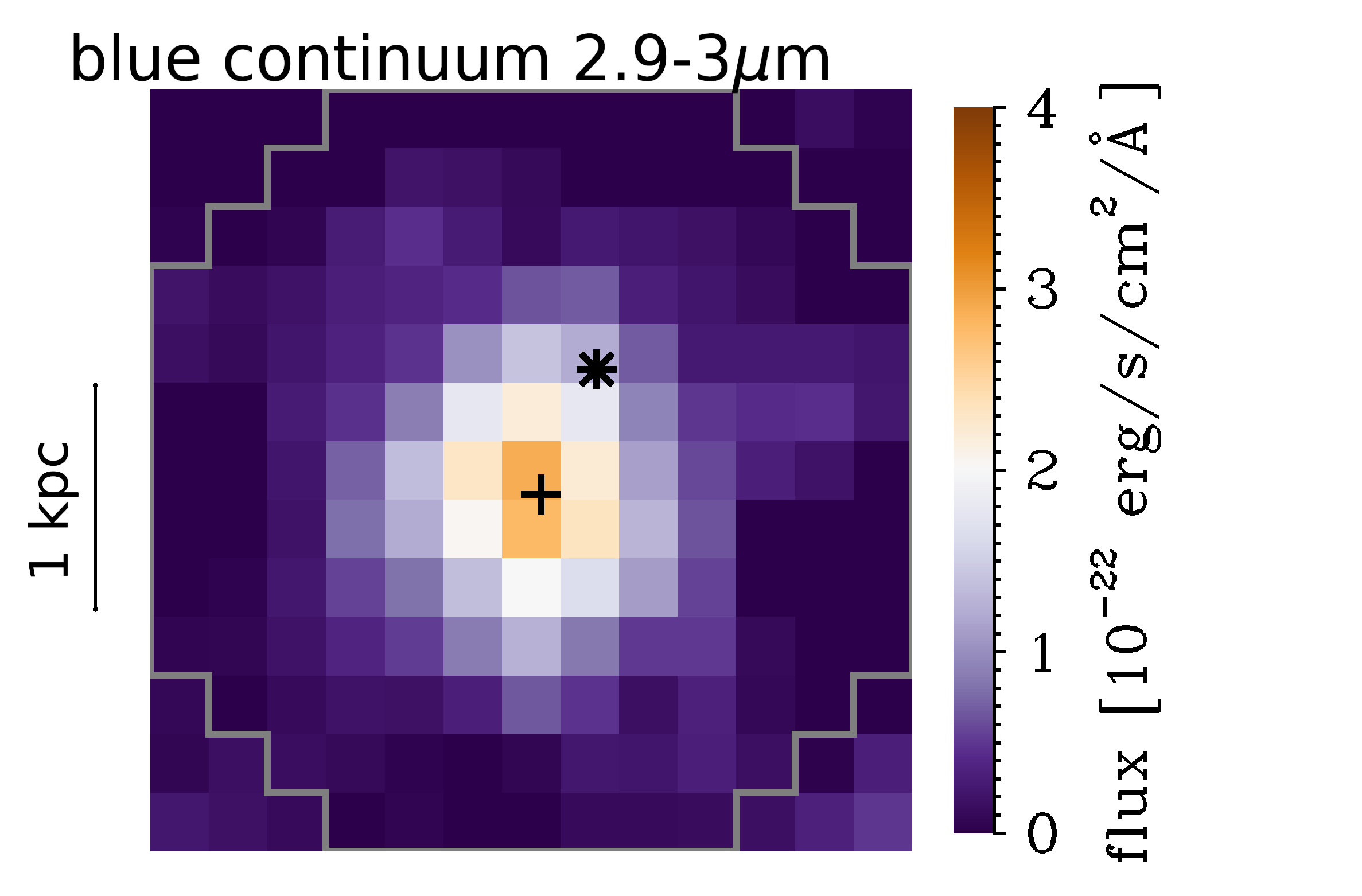}
    \includegraphics[width=0.3\textwidth]{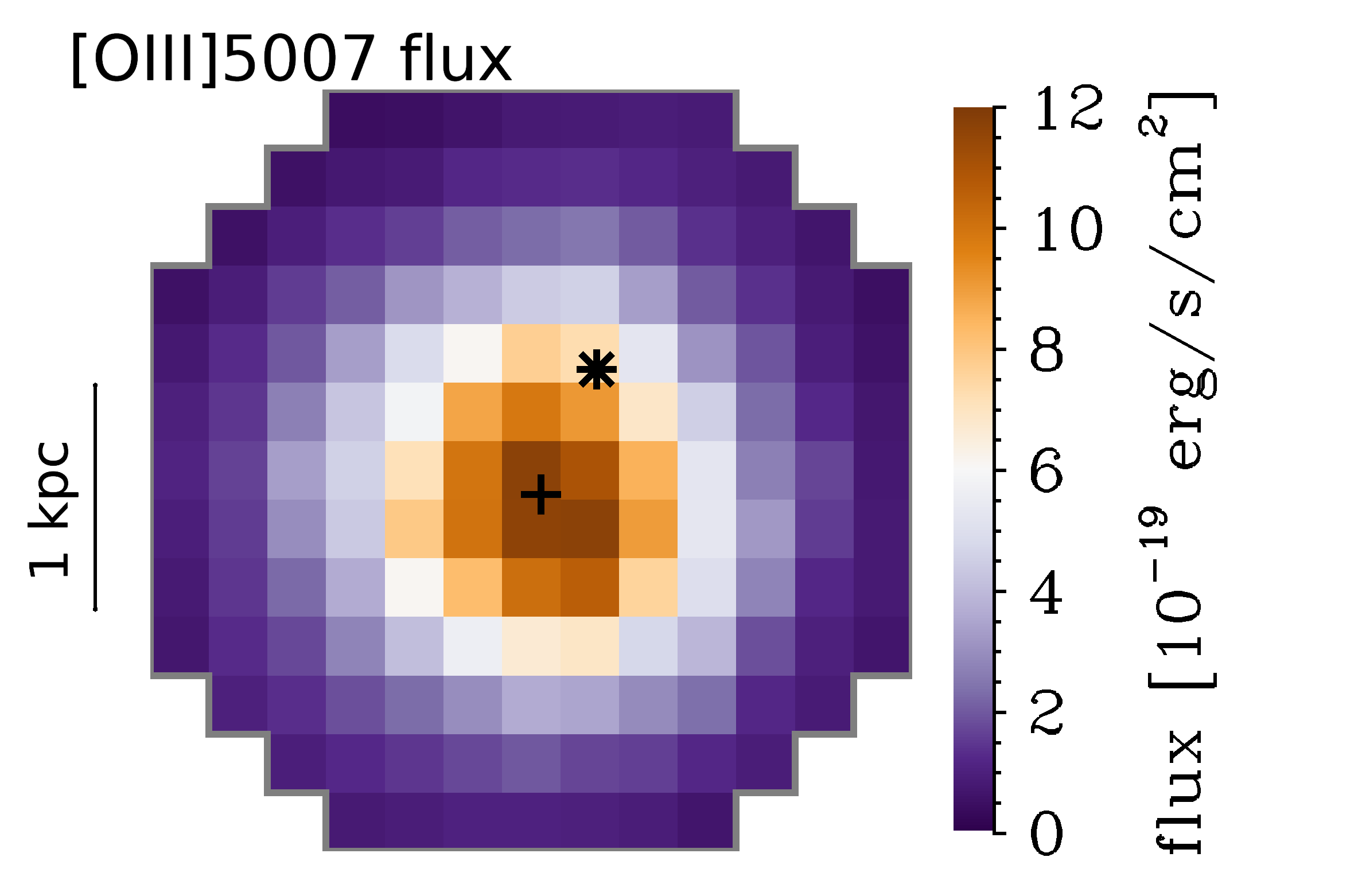}
    \includegraphics[width=0.3\textwidth]{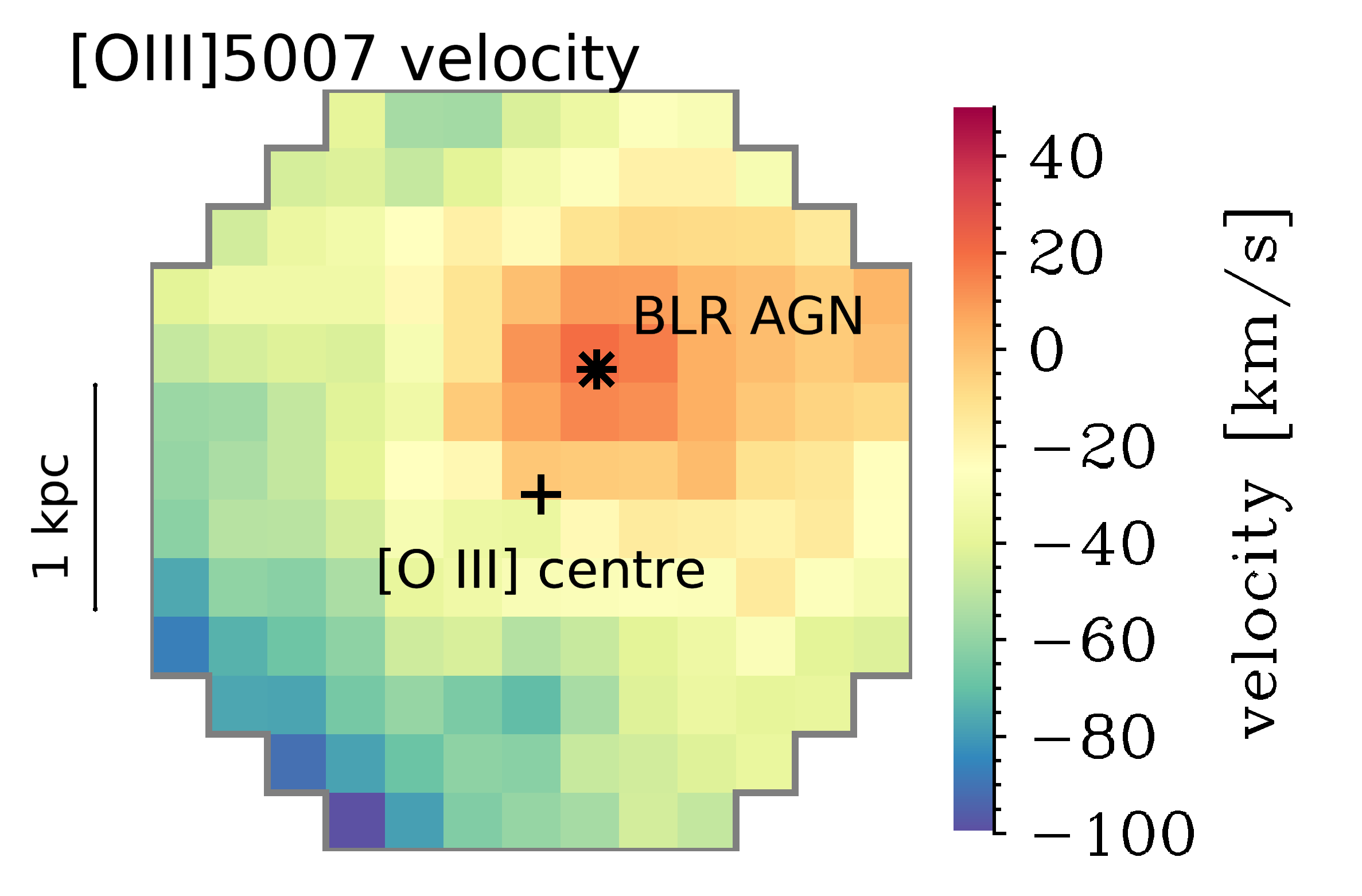}
    \includegraphics[width=0.3\textwidth]{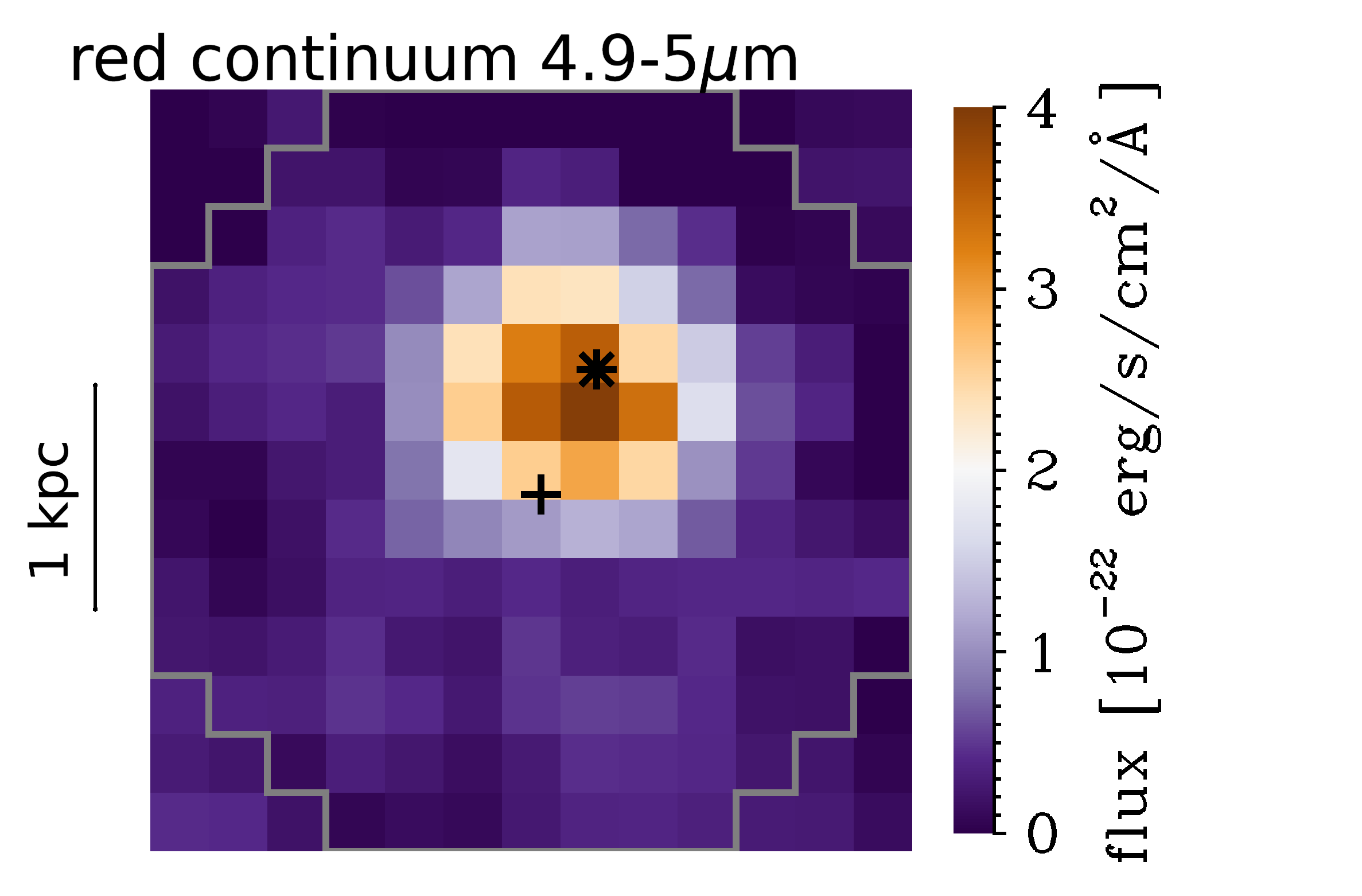}
    \includegraphics[width=0.3\textwidth]{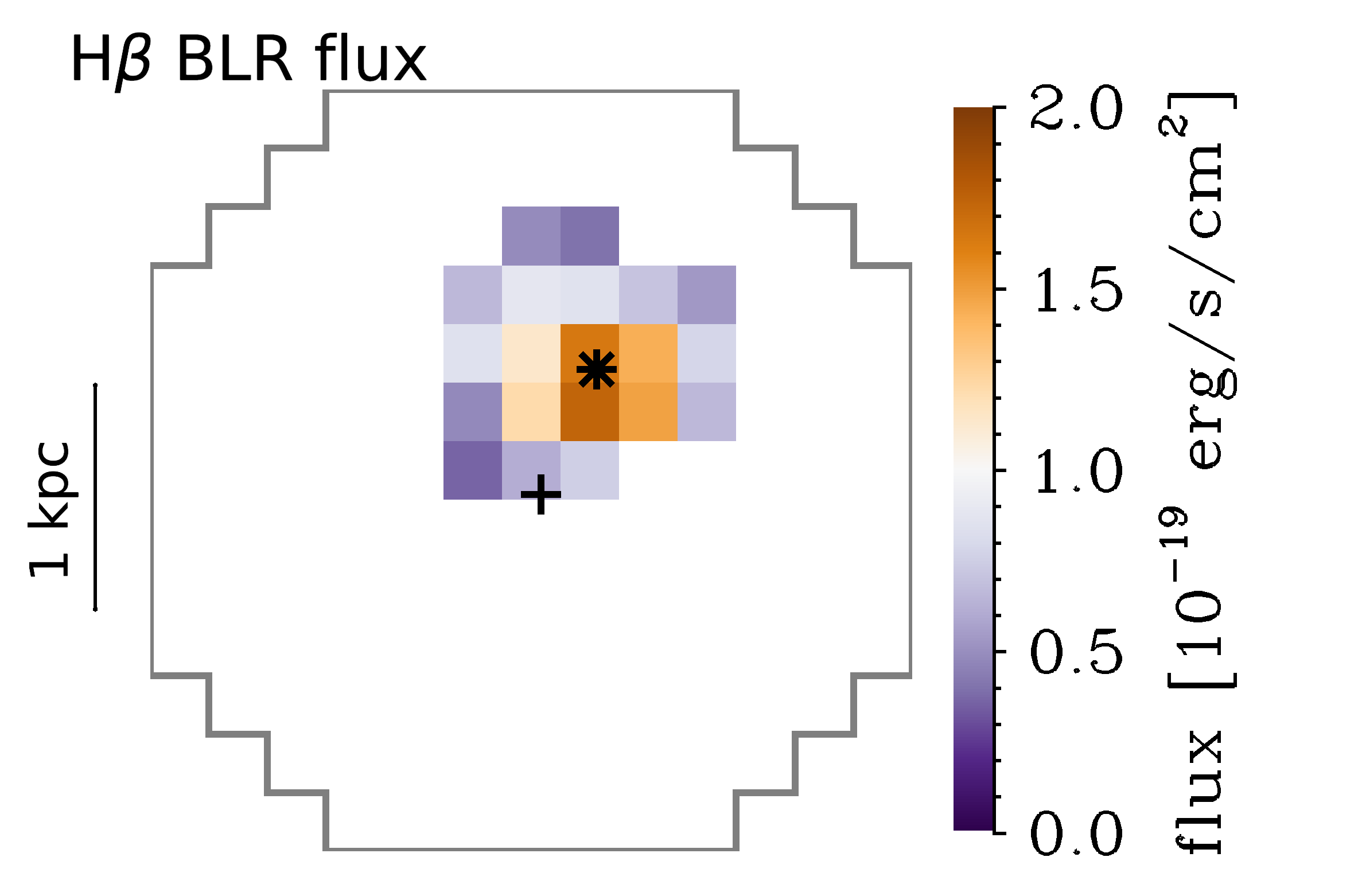}
    \includegraphics[width=0.3\textwidth]{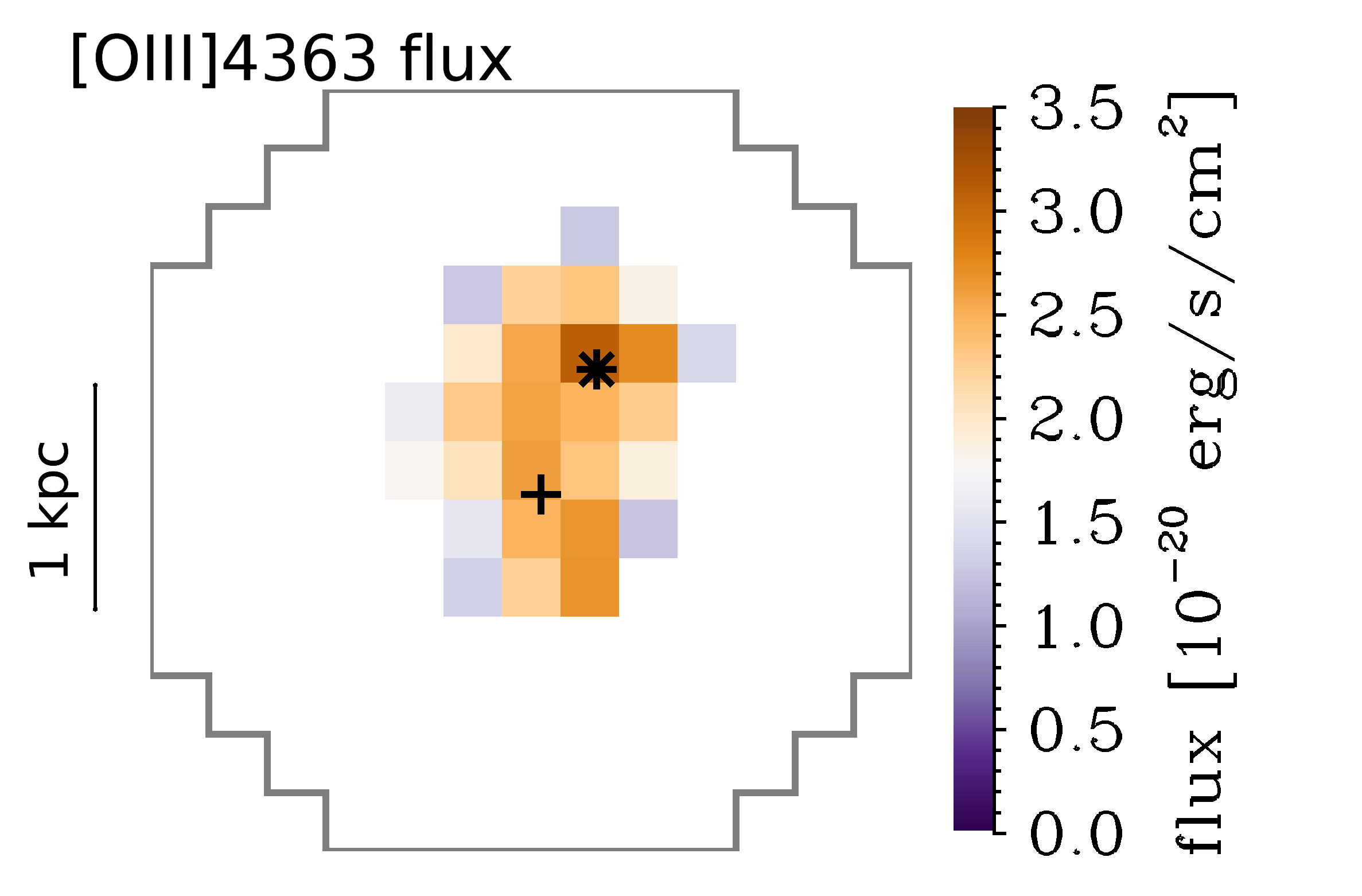}
    \caption{Top left: continuum flux density map at $\lambda=2.9-3\mu$m. Bottom left: continuum flux density map at $\lambda=4.9-5\mu$m.
    Top centre, bottom centre, and bottom right: integrated line flux maps with $S/N\geq5$ from our spectral fitting for the narrow \OIIIopL component, the \Hbeta BLR component, and the narrow \OIIIaL component, respectively. Top right: velocity map of the narrow \OIIIopall emission. 
    In each panel, the black cross and star indicate the centroid of the \OIIIopL emission and the BLR location as shown in Fig.~\ref{f:target}, respectively. The black scale bar to the left indicates $0.2''\sim1$~kpc, and the grey outline indicates the mask used for our full spectral fitting.
    There is a clear shift in the location of the continuum emission from bluer to redder wavelengths, aligning with the centroids of the \OIIIopL emission and the BLR position, reflecting the color gradient seen in the NIRCam image (Fig.~\ref{f:target}).
    The flux of the auroral \OIIIaL line peaks at the BLR position, but covers also the \OIIIopL flux peak, indicating a clear gradient of the \OIIIopL/\OIIIaL line ratio from the position of the AGN to the \OIIIopL emitter.}
    \label{f:maps}
\end{figure*}

We first fit the integrated spectrum extracted from $3\times3$ spaxels centred on the BLR location to robustly constrain the shape of the BLR emission. Measuring the BLR FWHM on a larger aperture including all spaxel with $S/N_{\rm H\beta,BLR}\geq5$ (see Fig.~\ref{f:maps}) gives consistent results within the uncertainties. Here and elsewhere, the $S/N$ is measured directly from the best fit. Then, we fit the emission extracted from individual spaxels by fixing the BLR FWHM to the previous best-fit value. We list the resulting emission line fluxes extracted at the BLR location and \OIIIopL centroid position in Table~\ref{t:fluxes}.
Zoom-ins on the corresponding spectral fits extracted from individual spaxels are shown in Fig.~\ref{f:target}.

If we integrate over a larger aperture centred on the \OIIIopL emission in ZS7, we can detect additional fainter lines. A circular aperture with radius $0.3^{\prime\prime}$ (encompassing both the \OIIIopL emitter and the BLR, similar to the region shown in Fig.~\ref{f:maps}) reveals tentative line emission at the expected wavelengths of H11, H10, H9, and a detection of He~I$\lambda5876$. 
While \HeIIL has been detected in previous observations from the ground \citep{Laporte17}, we do not detect \HeIIopL in ZS7. This is not necessarily surprising, because \HeIIL is the brighter line. If we extract a spectrum collapsed from a pseudo-slit corresponding to the previous X-Shooter observations by \cite{Laporte17}, we find an upper limit on the \HeIIopL flux which is consistent within the uncertainties with their measurement for \HeIIL within a factor of 8-13, consistent with theoretical expectations.

\subsection{Maps and velocity field}\label{s:maps}

We use the fluxes resulting from full spectral fitting to obtain emission line maps, for the regions where the lines are detected at more than 5$\sigma$. In Fig.\ref{f:maps} we show specifically the flux maps of \OIIIopL (which is also shown in Fig.\ref{f:target}), the broad \Hbeta component, and \OIIIaL.

The map of the broad component of \Hbeta is consistent with the PSF, i.e. a spatially unresolved source.
The maps confirm what was already visible from Fig.\ref{f:target}, namely the \Hbeta broad component is clearly offset from the \OIIIopL line centroid. 

The auroral line \OIIIaL peaks at the location of the \Hbeta broad, not unexpectedly, as strong \OIIIaL emission was already pointed out to be likely associated with AGN activity \citep{Dors2020,Brinchmann23,Perna17}. However it is interesting to see that it also extend to the \OIIIopL peak; we will discuss that the strength of the \OIIIaL line also at this location suggest the presence of an additional (type 2) AGN.

We also derive the \OIIIopL velocity map, as shown in Fig.~\ref{f:maps}. We use the same fitting approach as described in Section~\ref{s:fitting},but fit only the \OIIIopall emission which has the highest signal-to-noise ratio. The velocity is derived from the line centroid after subtracting the systemic velocity measured at the \OIIIopL centroid position. From our fits, we find an average uncertainty of 6~km/s on the velocities shown in Fig.~\ref{f:maps}.
The region around the BLR location appears clearly separated in velocity space, as indicated by the cluster of positive relative velocities. The absolute difference in velocity to the centroid of the \OIIIopL emitter is only about 20 km/s, but significant. 
Being at the the same redshift, the small projected distance and the small velocity difference suggest that we are witnessing the imminent merger of two systems, one of which hosts a BLR AGN.

Finally, we also extract and map the continuum in the bluest (2.9--3$\mu$m) and reddest (4.9--5$\mu$m) parts of the spectrum. As illustrated in Fig.~\ref{f:maps}, the red continuum peaks at the location of the broad \Hbeta, while the blue continuum peaks on the \OIIIopL centroid. This essentially confirms what was already seen in the NIRCam colors shown in Fig.~\ref{f:target}.
This further indicates that we are likely observing two distinct objects in close proximity.
The blue continuum centered on the \OIIIopL centroid is likely tracing a young stellar population. Instead, it is not straight-forward to interpret the nature of the compact red source at the location of the broad \Hbeta: it could be a reddened or aged stellar population, but it may also be associated with the reddened continuum of the AGN accretion disc. It is not possible to easily distinguish between these two scenarios, because the continuum spectral information is limited to 3600~\AA\ $<\lambda_{\rm rest}<5200$~\AA\ and has modest S/N.

\begin{table}
\caption{Observed narrow emission line fluxes extracted from one spaxel in units of $10^{-20}$~erg/s/cm$^2$ from full spectral fits to the BLR location and \OIIIopL centroid position and derived quantities.}\label{t:fluxes}%
\begin{tabular}{@{}lcc@{}}
\toprule
 & BLR location  & \OIIIopL centroid \\
\midrule
$[{\rm O~II}]\lambda\lambda3727,3729$  & $4.8\pm0.8$ & $5.8\pm0.6$\\
$[{\rm Ne~III}]\lambda3869$  & $4.6\pm0.5$ & $7.1\pm0.3$\\
${\rm He~I+H8}$  & $0.9\pm0.4$ & $2.7\pm0.3$\\
$[{\rm Ne~III}]\lambda3967$ + \Hepsilon  & $2.6\pm0.7$ & $3.9\pm0.5$\\
\Hdelta & $1.7\pm0.4$ & $4.1\pm0.3$ \\
\Hgamma & $2.9\pm0.4$ & $5.5\pm0.3$ \\
$[{\rm O~III}]\lambda4363$  & $3.1\pm0.4$ & $2.6\pm0.3$\\
\Hbeta & $6.2\pm0.5$ & $14.1\pm0.5$ \\
$[{\rm O~III}]\lambda5007$  & $72.9\pm1.8$ & $117.1\pm1.5$\\
\midrule
12 + log(O/H) & $7.69\pm0.08$ & $7.81\pm0.04$\\
$T_{e,\rm \OIIIopL}$ [K] & $22700\pm1900$ & $17400\pm500$\\
\midrule
\end{tabular}
\end{table}

\subsection{Stellar mass estimate for the \OIIIopL emitter}\label{s:mstar}

We estimate the stellar mass of the \OIIIopL emitter through a full SED modelling using imaging data from the PRIMER programme on the pixel level, following \cite{PerezGonzalez23} \citep[see also][]{2023ApJ...944....3G,DEugenio23}. SEDs were constructed for each one of the 262 30~mas (150~pc, with the PSF FWHM of the reddest image being around 5 pixels) pixels of the system counting with more then three NIRCam filters with $S/N>5$. The NIRCam data covers the wavelength range 0.9-4.4~$\mu$m with eight filters (one medium-band filter, the rest being broad-band). The SEDs were fitted to stellar population models by \cite{Bruzual03}, assuming a delayed exponential for the star formation history (SFH) with timescale $\tau$ taking values from 1~Myr (i.e., an instantaneous burst) to 10~Gyr (i.e., a constant SFH). The attenuation was treated with the \cite{Calzetti00} law, with A(V) values from 0~mag to 5~mag. The metallicity was left to vary between $0.005 Z_\odot$ and  $Z_\odot$. The IMF was set to \cite{Chabrier03}. With a total of four free parameters, the best fitting model (using a Monte Carlo method to estimate errors) is scaled to the photometry to get an estimation of the (surviving) stellar mass. 

To estimate the stellar mass of the \OIIIopL emitter, we exclude the Northern region where the BLR is detected with $S/N>5$ to mitigate contamination from the AGN (see Fig.~\ref{f:maps}). This leads to a total stellar mass of $M_\star \sim 2.9\pm0.9\times10^9$~M$_\odot$. If we fill in the excluded region with values from the Southern part of the galaxy at similar distance from the \OIIIopL centroid, we find $M_\star \sim 3.8\pm1.1\times10^9$~M$_\odot$. This estimate is consistent with what has been found in previous work from SED fitting to the CANDELS and also IRAC data \citep{Pentericci16, Laporte17}.
This shows that the \OIIIopL emitter is a moderately massive galaxy for its redshift \citep[see e.g.][]{Curti23b, Nakajima23}, expected to host an intermediate-mass to massive black hole in its centre.

In Section~\ref{s:type2} we will argue that the narrow line emission associated with \OIIIaL may also trace recent or ongoing AGN activity. Therefore, we make another estimate of the stellar mass of the \OIIIopL emitter by excluding not only the region where the H$\beta$ BLR is detected with $S/N>5$, but also the region where the narrow \OIIIaL is detected with $S/N>5$. This would lead to a stellar mass of $M_\star \sim 1.7\pm0.5\times10^9$~M$_\odot$ (without filling in the excluded regions with other values). We consider this a lower limit on the total stellar mass of the \OIIIopL emitter, and note that a reduction of the stellar mass by a factor 2--3 does not impact the results and conclusion presented in the following sections.

\section{Evidence for an offset type 1 AGN and environment of ZS7}\label{s:type1}

At a projected distance of 620~pc from the centroid of the \OIIIopL emission in ZS7, we detect a broad component at the wavelength of \Hbeta (Fig.\ref{f:target}). As discussed, the emission is consistent with stemming from a point source, spread by the point-spread function (PSF; PSF$_{\rm FWHM}\sim0.13-0.14''$ at 4~$\mu$m; see Fig.~\ref{f:maps}). The broad component does not have a counterpart in the forbidden lines, in particular not in the strong \OIIIopall doublet, hence it cannot be associated with emission of high velocity ISM or CGM, for instance due to galactic outflows. Therefore, the most plausible explanation is that the broad \Hbeta is tracing the Broad Line Region of an accreting black hole, i.e. a type 1 AGN (we discuss and discard other interpretations in Section~\ref{s:alternatives}).

The finding of an AGN in ZS7 is not so surprising, as evidence for an AGN was already found in the past based on the detection of \NVL and \HeIIL in the rest-frame UV \citep{Laporte17}. The novelty of our finding is that the type 1 AGN is offset relative to the \OIIIopL emitter. Also based on the kinematic offset around the BLR (Fig.S\ref{f:maps} and Sect.S\ref{s:maps}) and the markedly different colors, these results suggests that the type 1 AGN traced by the broad \Hbeta is hosted in a system separate from the galaxy traced by the bulk of the \OIIIopL emission, and in the process of merging with it. The ongoing merger may be channelling gas to the $5\times10^7$~M$_\odot$ black hole as a consequence of tidal interactions (\citealp{DiMatteo05, Johansson09}; see also e.g.\ \citealp{Dougherty24}). 

We consider the possibility that the BLR is actually at the centre of a single system but partially obscured by dust, while the \OIIIopL brighter emission region is tracing the less obscured part of the galaxy (or a less obscured part of an extended Narrow Line Region). However, the dust extinction towards nebular lines as traced by the \Hbeta/\Hgamma Balmer decrement does not show a significant gradient between the two locations and, if anything, slightly lower reddening towards the narrow nebular lines at the BLR location (see Appendix~\ref{a:ext} and Fig.~\ref{f:HgHb}). In addition, the fact the \OIIIaL line, which is clearly resolved across the two locations, peaks at the BLR location, provides further evidence that the BLR location is not the dust-obscured centre of the \OIIIopL emitter. In the case of differential reddening, one would expect that \OIIIaL would peak at the same location as \OIIIopL, whereas it peaks at the BLR location.

It is also implausible that the BLR is tracing the centre of a single system and that the \OIIIopL centroid is tracing a burst of star formation in its disc, at $\sim$600~pc from the nucleus. Indeed, the \OIIIopL emission is not an unresolved clump of star formation, it is a well resolved structure extending over $\sim 1.5$~kpc. Additionally, if the BLR was the centre of a single system then the velocity field should be approximately symmetric relative to it; on the contrary, the velocity field around the BLR stands out for having a clear offset relative to the surrounding region, just as expected for a separate system.
Furthermore, there is no evidence of significant extended line or continuum emission North-West of the BLR. Finally, we will show that also the \OIIIopL centroid is likely hosting another (type 2) AGN, indicating that it is associated with a separate nucleus.

At the same redshift as ZS7, we detect three fainter companions through emission lines of \Hbeta and \OIIIopall: two to the North at projected separations of 2.5 and 4.1 kpc, and one to the South-East at a projected distance of 3.9~kpc. Only the closest satellite is visible in the imaging data (see Figure~\ref{f:target}), and has been previously photometrically identified by \cite{Witten23b}.
The high number of companions, in addition to the offset BLR, suggests that we are witnessing a snapshot of the complex assembly of a larger system just 740 million years after the Big Bang.

\section{Black hole mass and luminosity of the offset type 1 AGN}\label{s:m_bh}

By assuming standard virial relations, we can estimate the black hole mass using our best-fit values of the \Hbeta BLR flux and FWHM, based on the 20-spaxel integrated spectrum around the BLR location, where $S/N_{\rm \Hbeta,BLR}\geq5$. The FWHM is corrected for the instrumental spectral resolution which is approximately $R\sim1000$ at the observed wavelength of \Hbeta, corresponding to $\sigma\sim127$~km/s \citep{Jakobsen22}.
Using the single-epoch calibration by \cite{Greene05} derived from data of nearby AGN, we find a black hole mass of $\log({\rm M_\bullet}/{\rm M_\odot})=7.7\pm0.4$. The uncertainties are dominated by the scatter of local scaling relations, which we take into account by adding 0.4~dex in quadrature to our error budget \citep[e.g.][]{Ho14}.
We also estimate the AGN bolometric luminosity to be $L_{\rm bol}=8.5(\pm0.8)\times10^{44}$~erg/s based on the \Hbeta BLR flux following \cite{DallaBonta20}, indicating that the black hole is accreting at about 14 per cent of its Eddington limit.
We report these measurements in Table~\ref{t:bh}.

\begin{table}
\caption{Measurements of the broad \Hbeta component extracted from an integrated spectrum with $S/N_{\rm \Hbeta,BLR}>5$ (see bottom centre panel of Fig.~\ref{f:maps}), and corresponding black hole properties.}
\begin{tabular}{lc}
\toprule
    Measurement & Value  \\
\hline 
    FWHM$_{\rm \Hbeta,broad}$ [km/s] & $3660\pm300$\\
    $F_{\rm \Hbeta,broad}$ [erg/s/cm$^2$] & $(1.7\pm0.2)\times10^{-18}$\\
    $L_{\rm \Hbeta,broad}$ [erg/s] & $(1.0\pm0.1)\times10^{42}$\\
    $\log(M_{\rm BH}/M_\odot)_{\rm \Hbeta,broad}$ & $7.7\pm0.4$ \\
    $\log(L_{\rm bol}/(\rm erg/s))_{\rm \Hbeta,broad}$ & $44.9\pm0.1$ \\
    $\log(L_{\rm Edd}/(\rm erg/s))$ & $45.8^{+0.7}_{-0.2}$ \\
    $\lambda_{\rm Edd}$ & $0.14$ \\
\end{tabular} 
\footnotesize{}
\label{t:bh}
\end{table}

\section{Excluding alternative scenarios for the broad \Hbeta: Supernovae, outflow and massive stars}\label{s:alternatives}

We briefly discuss alternative interpretations of the broad \Hbeta component: a local outflow, a supernova (SN) in the galaxy disc, or a cluster of Wolf-Rayet or very massive stars. 

While we recall that outflows as fast as a few 1000~km/s are only associated with AGN feedback (e.g.\ \citealp{Fabian12, Heckman14, Carniani15, FS19, Veilleux20}; see also discussion by \citealp{Maiolino23a}), the outflow scenario should result in even more prominent broad emission in \OIIIopall, as observed in galactic outflows locally and at high redshift, and especially because outflows are typically metal enriched with high \OIIIopL/\Hbeta \citep{Origlia04}.\footnote{We note that in high-metallicity systems (12+log(O/H)$\gtrsim8.5$) \OIIIopL can be suppressed with respect to \Hbeta \citep[e.g.][]{Curti20}. However, our metallicity estimates for ZS7 are much lower (see Table~\ref{t:fluxes}).}

Core collapse SNe do show broad Balmer line emission; at their peak luminosity their broad \Hbeta luminosity is of the order of $\sim 10^{40}$~erg/s, although some extremely luminous SNe can reach broad \Hbeta luminosities\footnote{We note that SN luminosities in the literature are generally reported for \Halpha, and \Halpha/\Hbeta can be quite large.}
of a few times $\sim 10^{41}$~erg/s \citep{Pastorello02,Tartaglia16,Kokubo19,Dickinson23,Fransson14}, still  lower than what is observed in ZS7 ($L_{\rm H\beta,broad}\sim 10^{42}$~erg/s; Table~\ref{t:bh}). The additional important difference is that when SNe reach such high luminosities, the broad component of \Halpha and \Hbeta is generally much larger than observed in ZS7, typically several 1000 km/s and even exceeding 10,000 km/s. Moreover, core-collapse SNe, especially the luminous ones, have an asymmetric blue-shifted profile (from a few 100 km/s to a few 1000 km/s) and/or P-Cygni profiles \citep[e.g.][]{GalYam12, Taddia13, Gutierrez17}, while the broad \Hbeta component in ZS7 is symmetric (Fig.~\ref{f:target}). Since we detect the continuum, we can also rule out strong absorption lines. We also note that the spectrum of core-collapse SNe (especially in the early phases) is further generally characterised by prominent \OIL emission and/or Fe bumps near \Hbeta, which are not seen in ZS7.

Additional compelling evidence against the SN scenario comes from the fact that in this case the source at the location of the broad \Hbeta should display some flux variation. We can leverage the fact that the NIRCam and NIRSpec-IFU observations were obtained at two different epochs separated by 5 months, i.e.\ 20 days rest-frame. While this is not a long time lag, depending on the phase of the putative SN light curve, we would expect a flux variation ranging from a factor of several to at least 20\%. We have consistently extracted the flux from the same aperture both from the NIRCam and NIRSpec data. For NIRCam we have used the filters overlapping with our NIRSpec spectrum, i.e. F444W, F410M and F356W, and in NIRSpec we have extracted the spectra matching the same wavelength range as the three filters. At least a 5\% cross-calibration uncertainty is estimated due to the fact that the PSF of the two instruments are not identical (i.e. slight differential losses, despite the large adopted aperture), as well as slight background disuniformities in the IFU data. Nonetheless we obtain that the flux in the three bands are fully consistent within 3\%, 2\% and 0.07\%, in the three band respectively. Such a good match, and the implied constancy of the flux (both of the continuum and of the broad \Hbeta, which is included in all three filters), makes the single SN scenario simply impossible.

However, one can explore whether the broad \Hbeta is not due to a single SN, but the overlap of multiple, co-located SNe associated with star formation within the small region associated with the BLR. However, even assuming the extreme case of no AGN contribution to the narrow \Hbeta, this would give an upper limit on the star formation rate of $< 6~M_\odot/$yr, i.e. $<10^{-1}~SN/$yr; even in the extreme case that the peak of the broad line emission of each SN lasts for one year, this would imply an average broad \Hbeta luminosity of $10^{39-40}$~erg/s, much lower than observed in ZS7. Also, one would expect the peak of the core collapse SNe to pile up at the location of the peak of star formation, i.e. at the \OIIIopL centroid.

Although it is not a common feature, broad Balmer lines have been associated with clusters of Wolf-Rayet stars in some individual cases \citep[e.g.\, in NGC 4214, see][]{Sargent91}. More distinctive features of Wolf-Rayet stars are for instance the `blue bump' around 4600-4720~\AA, associated with broad He~II$\lambda4686$, {\permittedEL[N][iii]}$\lambda4640$ and/or {\permittedEL[C][iii/iv]}$\lambda4650$, and the `red bump' around 5800~\AA, associated with {\permittedEL[C][iv]}$\lambda5808$ \citep[e.g.][]{Sargent91, Terlevich96, Schaerer97, Schaerer99, Brinchmann08a, MonrealIbero17}. These spectral signatures are not seen in ZS7. 
We calculate an upper limit on the flux in the `blue bump' region, and on He~II$\lambda4686$, from our aperture where the H$\beta$ BLR is detected with $S/N>5$. From this, we find `blue bump'/H$\beta<0.033$ and He~II$\lambda4686$/H$\beta<0.006$. For comparison, \cite{Sargent91} find `blue bump'/H$\beta\sim0.91$ in the region where they also detect broad H$\alpha$, and values of He~II$\lambda4686$/H$\beta\sim0.01-0.1$ are detected in nearby Wolf-Rayet galaxies \citep[e.g.][]{Vacca92, Thuan96, Fernandes04}. 
To further constrain from the H$\beta$ flux and the upper limit on He~II$\lambda4686$ the possible contribution from Wolf-Rayet stars to the spectral features in ZS7, we follow Equation~15 by \cite{Vacca92} to calculate the ratio of the number of WNL stars, the dominant sub-type of Wolf-Rayet stars, and the `equivalent' number of O stars, and find 0.019.
Considering the detection of He~II$\lambda1640$ in ZS7, the two independent observations discussed by \cite{Laporte17} reveal line widths that are consistent with the spectral resolution of the respective instruments, while again generally broad emission would be expected for Wolf-Rayet stars \citep[e.g.][]{Chisholm19, Leitherer19, Vanzella20}. 
In summary, while we cannot exclude the presence of Wolf-Rayet stars, we can conclude that their potential contribution to any spectral features in ZS7 is minor. This specifically holds for the broad H$\beta$ emission.

Very massive stars are also known to show broad Balmer lines, but in these cases the Equivalent Width of the broad component of \Hbeta is only a few \AA\ \citep{Martins2020}, while in ZS7 the Equivalent Width of the broad component of \Hbeta is about 60~\AA. Furthermore, in the case of very massive stars the broad \Hbeta is also accompanied by prominent and broad \HeIIopL \citep{Martins22}, not seen in ZS7.
Finally, we note that an outflow driven by a large numbers of massive stars has been invoked as a possible origin of broad off-nuclear \Halpha emission without a counterpart in \OIIIopall in a nearby galaxy \citep{RdP19}. However, in the latter case the width of the broad \Halpha is only $\sim 800-1000$~km/s.
Moreover, in the latter scenario, as well as all previous scenarios in which the broad \Hbeta is associated with clusters of young stars, one would expect the broad \Hbeta to peak at the location of the peak of star formation, i.e.\ at the location of the peak of (narrow) nebular emission, and not offset as in the case of ZS7.

Given the several arguments outlined above, and also the fact that previous studies already found evidence for an AGN in ZS7 \citep{Laporte17}, we conclude that the most plausible explanation is that the broad \Hbeta component is tracing the BLR of a massive, accreting black hole. Furthermore, the \Hbeta BLR and the \OIIIopL distribution are likely associated with two different systems in the process of merging. This was already discussed in Section \ref{s:type1}, and will be further discussed in the next two sections.

\begin{figure}
    \centering
    \includegraphics[width=\columnwidth]{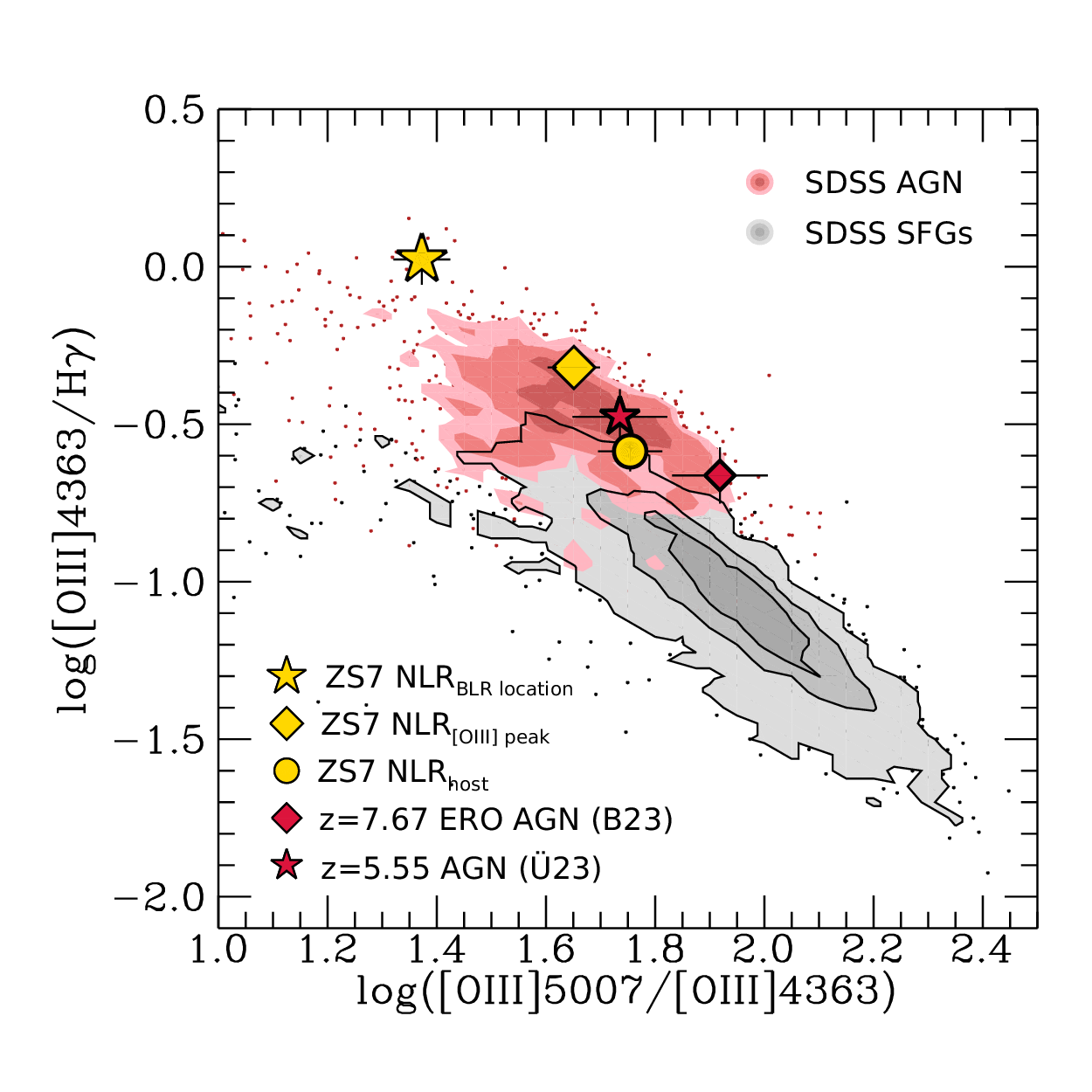}
    \caption{Narrow line ratio diagnostic diagram of \OIIIaL/\Hgamma\ {\it vs. }\OIIIopL/\OIIIaL. Different regions within ZS7 are indicated by golden symbols: the BLR location (star), the \OIIIopL centroid (diamond), and the surrounding region (circle; see Sec.~\ref{s:type2} for details). Local star-forming galaxies (grey shading and points) and AGN (red shading and points), dust-corrected following \citet{Cardelli89},  separate into two parallel sequences in this diagram, with AGN occupying a region of higher electron temperatures and having elevated ratios of \OIIIaL/\Hgamma. The red diamond shows a galaxy from the JWST Early Release Observations \citep{Pontoppidan22} at similar redshift, which is identified as an AGN through a [Ne~IV]$\lambda\lambda2422,2424$ detection (\citealp{Brinchmann23}; B23), and the red star indicates the narrow line ratios of a BLR AGN at $z=5.55$ (\citealp{Uebler23}; \"U23). The BLR centroid of ZS7 is located at the extreme end of the local AGN distribution, and also the \OIIIopL centroid falls on top of the local AGN cloud, while the host emission is located in a region where local star-forming galaxies and AGN overlap. The narrow line ratios extracted from the \OIIIopL centroid position could indicate recent AGN activity also in this part of ZS7.}
    \label{f:lineratios}
\end{figure}

\section{Indication of a second, type 2 AGN in the \OIIIopL emitter}\label{s:type2}

As discussed by various authors,  the detection of type 2 AGN, i.e. for which the BLR is obscured along our line of sight, and only the Narrow Line Region (NLR) is visible, is more challenging at high redshift. Indeed, the narrow line ratios of AGN at $z>4$ are displaced from the local AGN locus in classical optical diagnostic diagrams, such as the BPT \citep{Baldwin81}, involving \OIIIopL/\Hbeta vs [NII]6584/\Halpha, and the diagrams proposed by \cite{Veilleux87}, in which [NII]6584 is replaced by the [SII]6720 doublet \citep{Uebler23,Kocevski23,Harikane23,Maiolino23c,Scholtz23b}. This is likely a consequence of the gas in the NLR becoming metal-poor, which results in an overlap of AGN with the sequence of star-forming galaxies on these same diagrams.
Some of the alternative diagnostic diagrams to identify AGN based on UV lines proposed by, for instance, \cite{Scholtz23b}, are not usable in our case, due the limited spectral coverage of our data.

However, an additional diagnostic can be provided by the strong \OIIIaL auroral line, which is found to peak at the same location of the broad \Hbeta (Fig.~\ref{f:maps}), with an intensity comparable to \Hgamma. Such high \OIIIaL is extremely rare and, in combination with \OIIIopL, indicates very high gas temperatures, of the order of $T_{e, \rm \OIIIopL}=22700\pm1900$~K \citep[following][]{Dors2020}. 
Such a high ratio of \OIIIaL/\Hgamma is not seen in nearby star-forming galaxies, but it is seen in AGN \citep{Perna17, Brinchmann23}. 
This is further illustrated in Figure~\ref{f:lineratios}, where we plot nearby AGN (red contours) and star-forming galaxies (grey contours) from SDSS \citep{sdss7} in the \OIIIaL/\Hgamma\ {\it vs.\ }\OIIIopL/\OIIIaL line ratio diagram. In this diagram, local star-forming galaxies and AGN separate into two parallel sequences, with AGN occupying a region of higher electron temperatures and having elevated ratios of \OIIIaL/\Hgamma (Fig.~\ref{f:lineratios}; also Mazzolari et al., in prep.). 
Indeed, the narrow emission line ratios at the location of the broad \Hbeta in ZS7 lie at the extreme end of the nearby AGN distribution (golden star in Fig.~\ref{f:lineratios}). This is yet another confirmation that the broad \Hbeta is tracing a (type 1) AGN.

Intriguingly, as illustrated by the bottom right panel of Fig.~\ref{f:maps}, the auroral \OIIIaL line is spatially extended across the locations of both the BLR and \OIIIopL centroid. We can measure the \OIIIaL/\Hgamma and \OIIIopL/\OIIIaL line ratios at the \OIIIopL centroid position, and we find that also the emission line properties at the location of the \OIIIopL centroid are consistent with local AGN, as shown by golden diamond in Fig.~\ref{f:lineratios}. This finding strongly suggest that also the galaxy associated with the \OIIIopL centroid is hosting a (type 2) AGN.
 
We finally note that, if we extract these emission line ratios from an integrated spectrum excluding the region where the BLR is detected with $S/N>5$, and excluding the $3\times3$ spaxels around the \OIIIopL centroid, then we find values falling into the region which is locally occupied by both AGN and star-forming galaxies (golden circle in Fig.~\ref{f:lineratios}).

\section{An imminent galaxy and massive black hole merger}\label{s:merger}

As discussed in the previous Sections, the NIRSpec data and the NIRCam data provide evidence that ZS7 is composed of two components in close vicinity. 
The projected separation of 620~pc and velocity difference of the narrow line emission between the BLR location and the centroid of the \OIIIopL emitter ($\sim20$~km/s), likely hosting another AGN, are small, and likely imply an imminent merger.

In Section~\ref{s:m_bh}, we measured the mass of the black hole associated with the type 1 AGN, traced by the broad \Hbeta, by using virial relations. Estimating the mass of the putative second black hole, associated with the \OIIIopL emitter, is more difficult.
Using our stellar mass estimate of the \OIIIopL emitter ($M_\star\sim2.9\times10^9 M_\odot$, i.e. excluding the region where the BLR is significantly detected, see Section~\ref{s:mstar}) and assuming a $M_\star -M_\bullet$ relation calibrated on nearby AGN \citep{Reines15}, results in a secondary black hole mass associated with the \OIIIopL emitter of $\rm \log(M_\bullet/M_\odot)\sim5.8\pm0.6$. If we assume a lower limit for the stellar mass of $M_\star\sim1.7\times10^9 M_\odot$ (see discussion in Section~\ref{s:mstar}), we find $\rm \log(M_\bullet/M_\odot)\sim5.6\pm0.6$. Comparing this to the black hole mass estimate for the BLR AGN ($\log({\rm M_\bullet}/{\rm M_\odot})=7.7\pm0.4$), we find a ratio that is very similar to the dual AGN candidates identified by \cite{Maiolino23c}. 
If instead we assume a $M_\star -M_\bullet$ relation calibrated on local S/S0 and elliptical galaxies, we get a black holes mass higher by about one order of magnitude.
Alternatively, if we assume a black hole-to-stellar mass ratio between 0.01 and 0.1, as found for many high-z BLR-AGN discovered with JWST, the corresponding mass of the putative second black hole would be comparable to or even larger than the mass of the BLR-identified AGN in ZS7.
We note that simulations find that in ``offset'' AGN systems (i.e.\ interacting galaxies in which only one of the two black holes is accreting) the active black hole is typically more massive \citep{Steinborn16, Chen23}, although these results are only available at $z\lesssim3$.

Continuing with the interpretation of an imminent massive black hole merger, at a projected separation of 620~pc the black hole pair is likely in the dynamical friction phase when the interaction with stars and gas assists the black holes in reducing their angular momenta to sink to the common centre of mass  \citep{Chandrasekhar43, OstrikerE99}, before three-body interactions with individual stars \citep{Begelman80, Quinlan96} and eventually the emission of gravitational waves lead to coalescence of the black hole pair \citep{Peters63, Peters64}. 
Using our black hole mass estimate, the stellar mass estimate of the \OIIIopL emitter summed within roughly 3~kpc, and assuming an absolute separation of the BLR and the \OIIIopL centroid of 1~kpc, we find an approximative inspiral time of 100-200~Myr \citep{BT08}. 

\section{Implications}

It is not possible to provide any reliable statistics on the number and fraction of merging black holes at high redshift, based on this single case. We note, however, that this galaxy was observed as part of a sample of 11 galaxies at $z>6$ that were observed with the NIRSpec-IFU mode within the GA-NIFS survey. Of these, the ZS7 system and a second galaxy (Cresci et al., in prep.) are found to clearly host an AGN based on the detection of broad Balmer lines. 
The other AGN at $z>6$ in GA-NIFS is also found in an interacting system, on a few kpc separation. Also the two $z\sim6.8$ QSOs analysed so far within GA-NIFS undergo galaxy-galaxy mergers (\citealp{Marshall23}; see also e.g.\ \citealp{Yue21}).
Although these are low-number statistics, these results suggest that AGN (and more broadly black hole activity) in the early Universe are associated with galaxy interactions. In the case of ZS7, the separation of the individual systems is so small that the galaxies and their supermassive black holes will most likely merge within the next few hundred million years.
At slightly lower redshifts ($4<z<6$), a recent study by  \cite{Maiolino23c} identified a significant number (3/11) of candidate dual AGN based on the complex profile of broad Balmer lines in the NIRSpec multi-object JWST Deep Extragalactic Survey \citep[JADES;][without 2d spatial information]{Eisenstein23}. 
At $z\sim3$, \cite{Perna23b} identify 20-30\% dual AGN on separation scales of 3-30~kpc within GA-NIFS. 
Summarising, the findings of our and other studies using different approaches to assess black hole activity in the early Universe are consistent with a potentially large number of black holes in the process of merging.

There are no simulations available that provide the expected number of binary black holes, with separations less than 1~kpc, in this mass range and at $z\gtrsim7$. However, the mentioned finding of a large fraction of dual AGN at lower redshift (z$\sim$3) on scales of a few/several kpc already highlighted tension with the expectation of simulations, which predict a lower fraction of dual AGN at $z<5$ \citep{RosasGuevara19, DeRosa19, Volonteri22, Chen23} than observed \citep{Koss12, Spingola19, Perna23b}. 
However, while some simulations find an increasing fraction of massive black hole pairs with decreasing redshift due to long coalescence timescales of massive black hole mergers in these models \citep{Chen23}, most simulations predict the fraction of massive black hole pairs to increase with increasing redshift due to higher galaxy-galaxy merger rates (\citealp{Volonteri16, RosasGuevara19, DeRosa19, Volonteri22}; see also discussion by \citealp{Sesana09}).  

We also mention another possible scenario explaining our observations of ZS7, namely that the displaced accreting black hole actually results from the recoil of a recent supermassive black hole merger. An-isotropic emission of gravitational waves during a black hole merger event may give the resulting black hole a velocity kick \citep{Peres62, Bekenstein73}. Some simulations predict these velocities could be as high as several hundred to a few thousand km/s for pre-merger black holes with comparable masses, which could even lead to the ejection of the merged black hole from the host galaxy \citep[e.g.][]{Madau04, Merritt04, Blecha16}. Recoiling black holes may be observable through both a spatial and a velocity offset to the host galaxy \citep[see e.g.][]{Castignani22}, and this could be the case in ZS7: a spatial offset of 620~pc between the host galaxy emission and the BLR is evident in this case. In addition, we register a small velocity offset between the narrow line emission of ZS7 and the BLR emission of $\sim40$~km/s.  
While recoil velocities depend on the masses and relative spins of the initial black holes, the fact that we do not observe a large velocity offset of the BLR makes the interpretation of a recoiling supermassive black hole less appealing, although we caution that we cannot measure the absolute velocity difference.
Furthermore, a recoiled black hole would probably not have a NLR around it and, if it did, it would likely be the interstellar medium of the host galaxy, hence it should share the same velocity field as the host. In the case of ZS7 we clearly see that the narrow component of the nebular emission around the accreting black hole, i.e.\ its NLR, has a different velocity (Fig.~\ref{f:maps}). 
This suggests a more likely scenario in which the accreting black hole has yet to merge and is still embedded in the interstellar medium of its own galaxy. Finally, the indication that the \OIIIopL centroid is also likely to host another (type 2) AGN, implies that the merging has yet to happen, unless the narrow lines signatures are actually the fossil record of past accretion at that location. However, even the recoil scenario would be pointing at a black hole merging event; the difference between the two scenarios would be that in one case the merging has yet to happen, while in the other case it has just happened.

Finally, we note that another possible explanation for the offset BLR could be that it is tracing one of the so-called `wandering intermediate-mass black holes (IMBH)'. Recent simulations have shown that (at least on scales of several kpc) the dynamical friction during galaxy interactions/mergers may not be efficient at sinking black holes at high redshift \citep{DiMatteo23}. As a consequence, this may result into a potentially large population of IMBHs in the haloes of galaxies. However, such population of wandering IMBHs is expected to be on scales much larger (tens of kpc) than the separation observed in ZS7 (where dynamical friction is much more efficient). Additionally, a `wandering' black hole, even if accreting, should be essentially `naked', i.e.\ should not have its own NLR. This is in contrast with what is observed for the offset BLR in ZS7, which is clearly exciting at least \OIIIaL tracing low density gas outside the BLR.

\section{Conclusions}

The results presented in this work highlight the power of NIRSpec-IFU observations in the study of high$-z$ galaxies, mergers, and the detection of moderately massive black holes through imaging-spectroscopy.
Our observations provide clear and robust evidence for a massive black hole involved in a merger with another galaxy, likely hosting another accreting black hole, at $z=7.15$, only 740 Myr after the Big Bang. 
Overall, our results seem to support a scenario of an imminent massive black hole merger in the early Universe, highlighting this as an additional important channel for the early growth of black holes. Together with other recent findings in the literature, this suggests that massive black hole merging in the distant Universe is common.
Our observations may be used as guidance for the modelling of gravitational wave events originating from massive black hole mergers that will be detectable with future observatories like LISA \citep[e.g.][]{Haehnelt94, Jaffe03, Sesana05, Valiante21, LISA23}.


\section*{Acknowledgements}

We are grateful to the anonymous referee for a constructive report that helped to improve the quality of this manuscript.
We thank Martin Bourne, Massimo Dotti, Sophie Koudmani, Alberto Sesana, Debora Sijacki, Sandro Tacchella and Naoki Yoshida for useful discussions. 
AJB and GCJ acknowledge funding from the ``FirstGalaxies'' Advanced Grant from the European Research Council (ERC) under the European Union's Horizon 2020 research and innovation program (Grant agreement No.~789056).
BRdP, MP and SA acknowledge grant PID2021-127718NB-I00 funded by the Spanish Ministry of Science and Innovation/State Agency of Research (MICIN/AEI/ 10.13039/501100011033). 
FDE, JS and JW acknowledge support by the Science and Technology Facilities Council (STFC), by the ERC through Advanced Grant 695671 ``QUENCH'', and by the UKRI Frontier Research grant RISEandFALL.
H{\"U} gratefully acknowledges support by the Isaac Newton Trust and by the Kavli Foundation through a Newton-Kavli Junior Fellowship.
IL acknowledges support from PID2022-140483NB-C22 funded by AEI 10.13039/501100011033 and BDC 20221289 funded by MCIN by the Recovery, Transformation and Resilience Plan from the Spanish State, and by NextGenerationEU from the European Union through the Recovery and Resilience Facility.
JSD acknowledges the support of the Royal Society through a Royal Society Research Professorship.
MAM acknowledges the support of a National Research Council of Canada Plaskett Fellowship, and the Australian Research Council Centre of Excellence for All Sky Astrophysics in 3 Dimensions (ASTRO 3D), through project number CE170100013.
PGP-G acknowledges support from grant PID2022-139567NB-I00 funded by Spanish Ministerio de Ciencia e Innovación MCIN/AEI/10.13039/501100011033, FEDER, UE.
RM acknowledges support by the Science and Technology Facilities Council (STFC), from the ERC Advanced Grant 695671 ``QUENCH'', and funding from a research professorship from the Royal Society.
SCa and GV acknowledge support from the European Union (ERC, ``WINGS'', 101040227).

\section*{Data Availability}
The NIRSpec data used in this research has been obtained within the NIRSpec-IFU GTO programme GA-NIFS and will become publicly available in May 2024. The NIRCam observations of ZS7 are public through the PRIMER programme (PI: James Dunlop). Data presented in this paper will be shared upon reasonable request to the corresponding author.


\bibliographystyle{mnras}
\bibliography{literature}

\begin{appendix}

\section{Full spectral fits}\label{a:fits}

\begin{figure*}
    \centering
    \includegraphics[width=\textwidth]{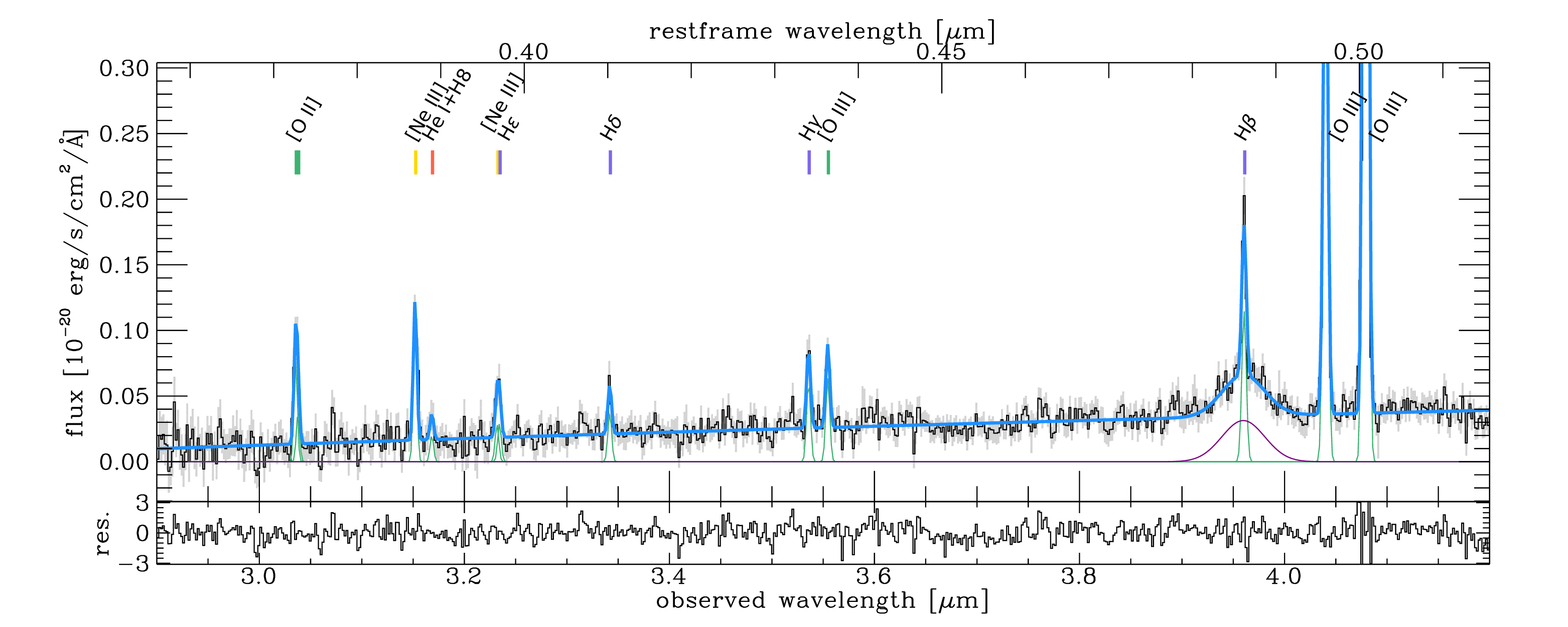}
    \includegraphics[width=\textwidth]{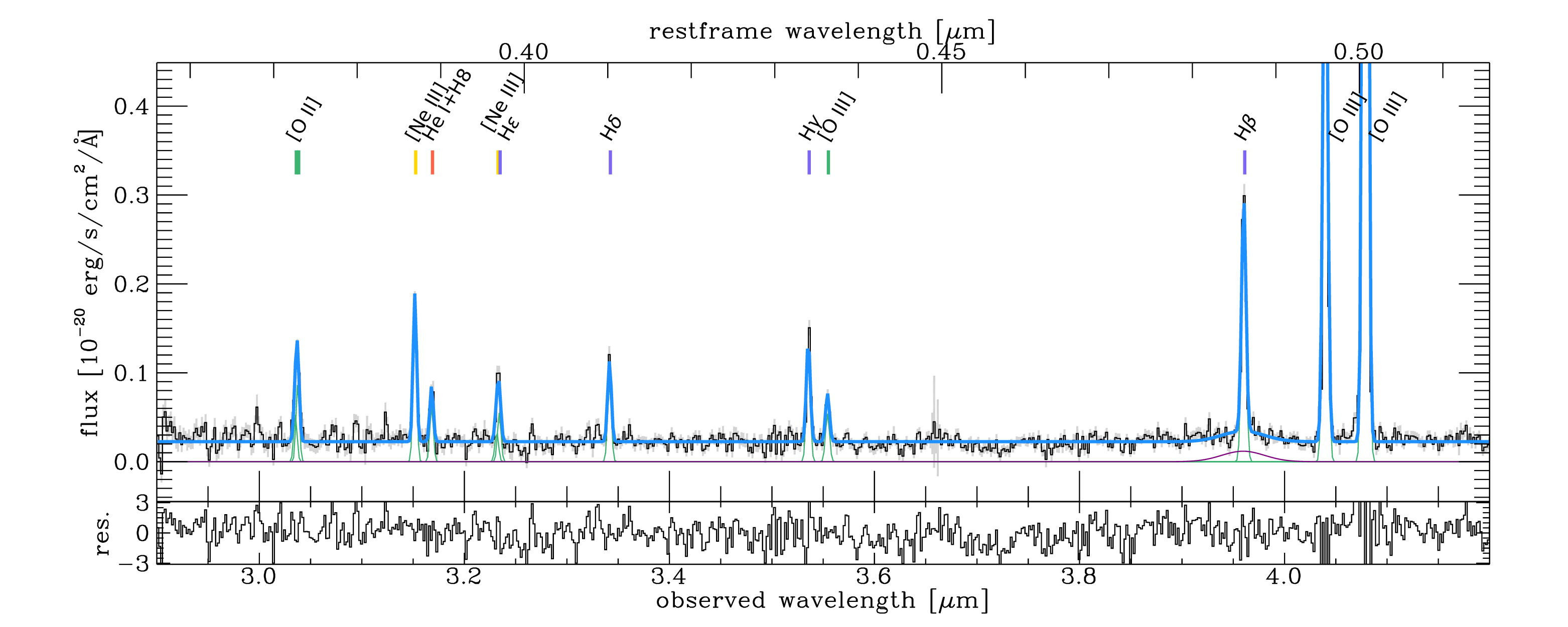}
    \caption{Full spectral fits to spectra extracted from individual spaxels at the BLR location (top), and at the centroid location of the \OIIIopL emitter (bottom). Residuals show (data - model)/uncertainties.
    We show our full fit in blue, with narrow emission lines in green and the BLR component in purple. 
    At the BLR location, the broad \Hbeta emission is evident, while the \OIIIopall emission is narrow at all locations. We note the strong auroral \OIIIaL emission line in both spectra.}
    \label{f:fullfits}
\end{figure*}

In Fig.~\ref{f:fullfits} we show full spectral fits to spectra extracted from individual spaxels at the BLR location, and at the centroid location of the \OIIIopL emitter. At the BLR location, the broad \Hbeta emission is evident. Some flux of the BLR point-source is spread through the point-spread function (PSF$_{\rm FWHM}\sim0.13-0.14''$ at $4\mu$m) and therefore is still visible at the \OIIIopL centroid position.

\section{Dust extinction}\label{a:ext}

\begin{figure}
    \centering
    \includegraphics[width=0.6\columnwidth]{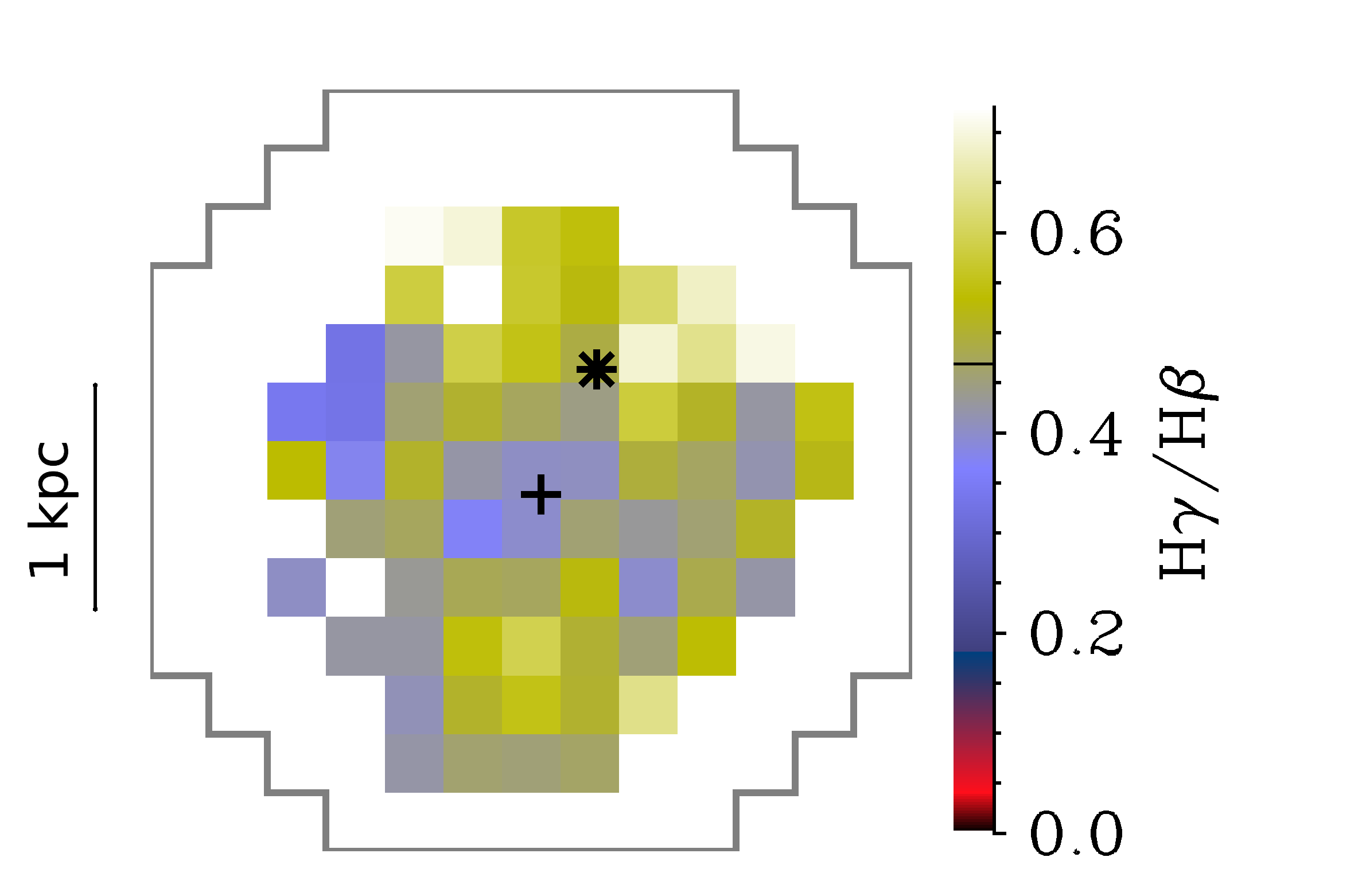}
    \caption{Map of the PSF-matched \Hgamma to \Hbeta narrow emission line flux where both lines have $S/N>5$, with the centroid location of the \OIIIopL and the location of the BLR indicated by a cross and star, respectively.}
    \label{f:HgHb}
\end{figure}

In Fig.~\ref{f:HgHb} we show a map of the PSF-matched \Hgamma to \Hbeta narrow emission line flux, derived using the empirical fitting functions for the NIRSpec-IFU PSF provided by \cite{DEugenio23}. The centroid locations of the \OIIIopL and BLR are indicated by a cross and star, respectively. The black line in the colorbar shows the theoretical line ratio of \Hgamma/\Hbeta=0.468, whereas lower values indicate dust extinction. The data show that there is no enhanced dust extinction at the location of the BLR. This rules out a scenario where the BLR centroid location is the true, dust-obscured centre of the ZS7 system.

\end{appendix}


\bsp	
\label{lastpage}
\end{document}